\documentclass[conference]{IEEEtran}
\usepackage{rotating}

\IEEEoverridecommandlockouts
\usepackage{cite,xurl}
\usepackage{multirow}
\usepackage{amsmath,amssymb,amsfonts}
\usepackage{algorithmic}
\usepackage{graphicx}
\usepackage{textcomp}
\usepackage{algorithm2e}
\usepackage{xcolor}
\def\BibTeX{{\rm B\kern-.05em{\sc i\kern-.025em b}\kern-.08em
    T\kern-.1667em\lower.7ex\hbox{E}\kern-.125emX}}

\usepackage{caption}
\usepackage{subcaption}
    
\usepackage{tikz}
\usetikzlibrary{quantikz}

\usepackage{subcaption}

\usepackage{amsmath}
\usepackage{amssymb}
\usepackage{amsthm}

\newtheoremstyle{ieeestyle}
  {\topsep}   
  {\topsep}   
  {\itshape}  
  {}          
  {\bfseries} 
  {.}         
  { }         
  {}          

\theoremstyle{ieeestyle}

\begin{document}

\title{Encoding and Node Choices in Transversal Fault-Tolerant Distributed  Quantum Computations: An Initial Study}

\author{\IEEEauthorblockN{Seng W. Loke}
\IEEEauthorblockA{\textit{School of Information Technology, Deakin University, Burwood, VIC 3125, Australia.} \\
seng.loke@deakin.edu.au}
 }


\maketitle

\begin{abstract}
We compare and study different Bivariate-Bicycle (BB) encodings and node choices for distributed quantum operations such as transversal non-local CNOTs. We observe that while some encodings have more physical qubits requiring more ebits for a distributed computation,  reducing ebit consumption alone might not be the best criterion for selecting an encoding for the logical qubits, if the goal is to reduce the logical error rate of the distributed computation; for example, one should also consider encodings with a larger distance, which using more physical qubits can  provide. We consider distributed, or non-local, CNOTs and computation of the global gate (GCZ) over distributed logical qubits as examples. The choice of encoding enables particular concurrent operations; e.g., we show that a transversal physical GCZ on a self-dual $[[120,8,12]] BB$ code can realize eight concurrent logical GCZ operations after accounting for the logical permutation induced by transversal Hadamard.
\end{abstract}

\begin{IEEEkeywords}
distributed quantum computing, transversal gate gadgets, fault-tolerant quantum computing,  distributed  gates
\end{IEEEkeywords}

\section{Introduction}

Distributed or modular quantum computing can help  scale up quantum computation, via combining quantum and classical connected nodes (or QPUs)~\cite{BARRAL2025100747,knörzer2025,CALEFFI2024110672,arquin}.\footnote{For example, see ~\cite{photonics}, \url{https://spectrum.ieee.org/quantum-computers}, \url{https://newsroom.ibm.com/2025-11-20-ibm-and-cisco-announce-plans-to-build-a-network-of-large-scale,-fault-tolerant-quantum-computers}. See also \url{https://quantnet.lbl.gov} and \url{https://www.ox.ac.uk/news/2025-02-06-first-distributed-quantum-algorithm-brings-quantum-supercomputers-closer}.} 
For fault-tolerant distributed quantum computing, a range of quantum error correcting encodings  have been proposed   and  considered for distributed quantum computing, such as  surface codes (e.g., ~\cite{PhysRevLett.104.180503,singh2025modular,Chandra:2026ovs,ds45-fm9n,Rennela:2026aru}), Bivariate-Bicycle (BB) codes~\cite{Stack2026Transversal}, and floquet codes~\cite{sutcliffe2025distributed}, with different architectural approaches~\cite{11410221}.  
Previous work also looked at distributed transversal operations for CNOT using the Bivariate-Bicycle (BB)  encoding, e.g.,~\cite{Stack2026Transversal}, and a modular architecture using BB codes~\cite{yoder2025tour}.
Different distributed realizations of the
$[[144,12,12]]BB$   code block over 4, 6, and 12 QPUs were investigated in~\cite{Chandra:2026ovs}.
Here, we consider encoding and node choices, e.g.,   instead of distributing a block over multiple nodes,  consider a node hosting multiple blocks (if there is any benefit in this), the encoding for each block on each node, and so on.


For BB encodings, some common BB codes we consider in this paper include $[[36,4,6]]$, $[[54,4,8]]$, $[[90,8,10]]$, and $[[144,12,12]]$, in the form $[[n,k,d]]$, with $n$ physical qubits encoding $k$ logical qubits with distance $d$. Normally, the smaller  $n$ is the better, for a given $k$ to achieve  a given $d$, for encoding on a single QPU device, and higher $n/k$ can yield higher $d$. For transversal operations with encoding $[[n,k,d]]$, $k$ logical operations are enacted via $n$ physical operations, for the BB codes studied here. In a distributed setting, using blocks with large $n$ providing larger $d$, with  distributed inter-node  transversal operations requiring physical ebits (and  communication qubits and perhaps ancilla qubits on each node) corresponding to each physical qubit, a large $n$ might   mean many more inter-node physical ebits required, perhaps reducing the advantage of the larger  distance.

We consider a setting where there are circuits to be executed and there is a choice of nodes available for hosting and executing the circuits (e.g., {\em assuming} a cloud service model with a provider providing distributed QPUs as a service). Also, suppose that the set of  nodes (or QPUs) to be selected and used can be chosen from a heterogeneous collection (e.g., nodes have different numbers of qubits, and different possible encoding choices,\footnote{Although not on BB codes, earlier work demonstrated code switching and the ability to change QEC encodings on the same hardware~\cite{Pogorelov2025Experimental}.} etc). In this paper, we look into two questions: 
 
\begin{itemize}
    \item Which encoding to use for the logical qubits? For example, suppose a circuit has 16 (logical) qubits $q_1,...,q_{16}$, and there are two nodes available with enough qubits to allow   each of the two nodes to host 8 logical qubits, i.e., a $[[90,8,10]]BB$ block on each node; now, suppose part of the computation requires inter-node CNOTs between qubits $q_i$ and $q_{8+i}$ (for $i \in \{1,...,8\}$), which can be performed using a (non-local) transversal CNOT~\cite{Stack2026Transversal} between the two blocks, requiring 90 physical ebits; but  an alternative is to consider each node using two $[[36,4,6]]BB$ blocks (blk1 and blk2 say), and then the inter-node CNOTs can be done via two transversal CNOT operations between blk1 of node1 and blk1 of node 2, and between blk2 of node1 and blk2 on node2, say. Now,  we have reduced the distance to 6   compared to 10 in the one block encoding $[[90,8,10]]$, but would require only 72 ebits, i.e. 36 ebits for the physical CNOTs between the physical qubits from each block; hence, there seems to be a trade-off between distance and ebit consumption - how this affects the logical error rate (LER) is a question. Two smaller blocks on each node might also allow easier inter-node logical operations involving subsets of the qubits on both nodes (e.g., an operation with many interactions between qubits in blk1 of node1 and qubits in blk1 of node2, and none between qubits in blk2 of node1 and qubits in blk2 of node2). 
    \item Which combination of nodes to use for the circuit? For  example, for a circuit requiring $16$ logical qubits say, and suppose there is no node with enough physical qubits to host all $16$ logical qubits, but  suppose what is available are two nodes which can host one block of $[[90,8,10]]$ each, or  two nodes which can host one block of $[[36,4,6]]$ on one and three blocks of $[[36,4,6]]$ on the other.
\end{itemize} 

This paper discusses the above issues and provides an analysis in an attempt to answer the above questions. In the paper, we assume nodes have data or {\em computation qubits} for the actual computations, including {\em ancilla qubits}, and {\em communication qubits} for hosting ebits.
We examine the encoding and node choices in a number of scenarios below, first for the transversal CNOT in Section~\ref{sec:cnot}, and then for a distributed global gate (GCZ) in Section \ref{sec:gcz}. We discuss results in Section~\ref{sec:dis}, and then conclude in Section~\ref{sec:conc}.

\section{Encoding and Node Choices for the Non-Local Transversal CNOT} 
\label{sec:cnot}

We first consider choices of encodings and nodes for 16 logical qubits involved in a simple circuit (which may be part of some larger circuit) performing an inter-node transversal CNOT between the qubits on the two nodes~\cite{Stack2026Transversal}. We assume that the qubits have been allocated to the nodes $A$ and $B$ based on some   algorithm (e.g., there are many qubit allocation algorithms to reduce ebits between qubits across nodes), but still requires a transversal CNOT between them - this can happen for example, when there are a lot more interactions between qubits allocated to the same node, but an  inter-node transversal CNOT is still part of the computation (i.e., part of a larger circuit). Then, once the 16 (logical) qubits have been allocated to the two nodes (say equal number on each node) according to some criteria, we can then consider how the logical qubits could be encoded on the nodes; e.g., we can use (i) one $[[90,8,10]]BB$ block per node, or (ii) two $[[36,4,6]]BB$ blocks per node. 
We compare the LERs for a transversal CNOT, denoted by  $\overline{CNOT}_{A,B}$, between two   blocks   $A$ and $B$.

\begin{figure}
\centering
  \scalebox{0.75}{
      \begin{quantikz}
   \lstick[2]{A}   \ket{c}   \qw   & \qw   & \ctrl{1}  & \qw   & \qw & \qw & \qw &          \qw &  \qw   &   \gate{Z} & \qw \ket{c} \\
    c_A=\ket{0}     \qw & \ctrl{}{}  & \targ{}   & \qw & \meter{} &  \cwbend{1} &          &            &  &         \\
    \lstick[2]{B}    c_B=\ket{0}    \qw & \ctrl{}{}  & \qw       & \qw & \qw      & \gate{X}    & \ctrl{1}  & \gate{H}  & \meter{}  &  \cwbend{-2}       \\
       \ket{t}   \qw & \qw        & \qw       & \qw & \qw      & \qw         & \gate{U}       & \qw        & \qw &  \qw &\qw \ket{t'}
         \arrow[from=3-2,to=2-2,squiggly,dash,line width=0.1mm]{}
    \end{quantikz}   
    }
  \caption{Distributed controlled-$U$ (i.e., dCNOT is when U=$X$)   between nodes $QC$ and $QC'$ (the {\em computation} qubits are $c$, the control qubit, and $t$, the target qubit), and the wavy line illustrates a Bell pair  involving  the  {\em communication} qubits from the two nodes both initially $\ket{0}$. }
\label{dcnot}
\end{figure}
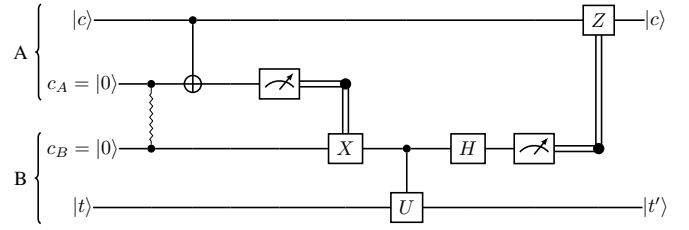

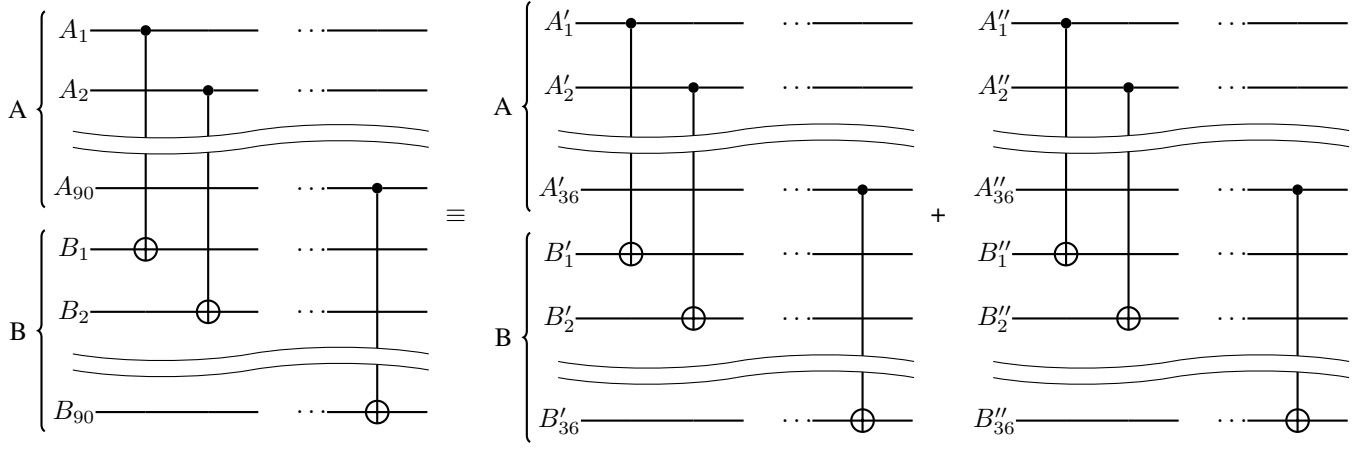
\begin{figure*}
\begin{center}

\begin{quantikz}
 \lstick[4]{A}  A_1    & \ctrl{4} & \qw      & \qw      & \ldots & \qw      & \qw \\
 A_2    & \qw      & \ctrl{4} & \qw      & \ldots & \qw      & \qw \\
 \wave  &          &          &          &         &          &     \\
 A_{90} & \qw      & \qw      & \qw      & \ldots & \ctrl{4} & \qw \\
\lstick[4]{B} B_1    & \targ{}  & \qw      & \qw      & \ldots & \qw      & \qw \\
 B_2    & \qw      & \targ{} & \qw      & \ldots & \qw      & \qw \\
 \wave  &          &          &          &         &          &     \\
 B_{90} & \qw      & \qw      & \qw      & \ldots & \targ{} & \qw
\end{quantikz}
    ~$\equiv$~
    \begin{quantikz}
 \lstick[4]{A}  A'_1    & \ctrl{4} & \qw      & \qw      & \ldots & \qw      & \qw \\
 A'_2    & \qw      & \ctrl{4} & \qw      & \ldots & \qw      & \qw \\
 \wave  &          &          &          &         &          &     \\
 A'_{36} & \qw      & \qw      & \qw      & \ldots & \ctrl{4} & \qw \\
\lstick[4]{B} B'_1    & \targ{}  & \qw      & \qw      & \ldots & \qw      & \qw \\
 B'_2    & \qw      & \targ{} & \qw      & \ldots & \qw      & \qw \\
 \wave  &          &          &          &         &          &     \\
 B'_{36} & \qw      & \qw      & \qw      & \ldots & \targ{} & \qw
\end{quantikz} 
~+~
\begin{quantikz}
   A''_1    & \ctrl{4} & \qw      & \qw      & \ldots & \qw      & \qw \\
 A''_2    & \qw      & \ctrl{4} & \qw      & \ldots & \qw      & \qw \\
 \wave  &          &          &          &         &          &     \\
 A''_{36} & \qw      & \qw      & \qw      & \ldots & \ctrl{4} & \qw \\
  B''_1    & \targ{}  & \qw      & \qw      & \ldots & \qw      & \qw \\
 B''_2    & \qw      & \targ{} & \qw      & \ldots & \qw      & \qw \\
 \wave  &          &          &          &         &          &     \\
 B''_{36} & \qw      & \qw      & \qw      & \ldots & \targ{} & \qw
\end{quantikz}
    \end{center}
    \caption{Transversal $\overline{CNOT}$ using $[[90,8,10]]$ blocks, one on each node, and using two $[[36,4,6]]$ blocks on each node. }
    \label{fig:trans-cnot}
    \end{figure*}

With (i), we perform 90 inter-node physical CNOTs, each denoted by $CNOT$, between corresponding physical qubits on the two nodes (nodes $A$ and $B$, but when only one block per node, we simply call the blocks $A$ and $B$ as well),  i.e. we perform:
\[
\overline{CNOT}_{A,B} =\bigotimes_{i=1}^{90} CNOT[{A_{i} \rightarrow B_{i}}]
\]
where we denote the $i$-th physical qubit on node $A$ by $A_{i}$ and the corresponding $i$-th physical qubit on node $B$ by $B_{i}$. This is illustrated on the left hand side in Figure~\ref{fig:trans-cnot}.
Each distributed CNOT gate (a.k.a. non-local CNOT) can be performed using the gadget  as shown in Figure~\ref{dcnot} from~\cite{eisert}. 

 Note that this performs 8 inter-node logical CNOTS, each denoted by $\overline{CNOT}_L$, concurrently, between the logical qubits encoded by the physical qubits:
\[
\bigotimes_{l=1}^{8} \overline{CNOT}_L[A^{(l)} \rightarrow B^{(l)}]
\]
where we denote the $l$-th logical qubit on node $A$ by $A^{(l)}$ and the corresponding $l$-th logical qubit on node $B$ by $B^{(l)}$.

With (ii) we perform 36 inter-node physical CNOTs between the qubits across the nodes, for each pair of blocks across the nodes. The two blocks on node $A$ are denoted by $A'$ and $A''$ and those on $B$ by $B'$ and $B''$. So we perform:
\[
\overline{CNOT}_{A',B'} = \bigotimes_{i=1}^{36} CNOT[{A'_{i} \rightarrow B'_{i}}]
\]
and
\[
\overline{CNOT}_{A'',B''} = \bigotimes_{i=1}^{36} CNOT[{A''_{i} \rightarrow B''_{i}}]
\]
which results in the inter-node logical $\overline{CNOT}_L$s:
$\bigotimes_{l=1}^{4} \overline{CNOT}_L[A'^{(l)} \rightarrow B'^{(l)}]$
and $\bigotimes_{l=1}^{4} \overline{CNOT}_L[A''^{(l)} \rightarrow B''^{(l)}]$. This is shown on the right hand side of Figure~\ref{fig:trans-cnot}.

  Both  (i) and (ii) can be interpreted as equivalent in terms of the inter-node operations between the 8 pairs of logical qubits. (i) requires 90 (physical) ebits for the 90 non-local physical CNOTs using the gadget in~\cite{eisert}, while (ii) requires 2x36=72 ebits in total for the 72 non-local physical CNOTs. Hence, there is a reduction of 18 ebits required, which may reduce errors potentially, at the cost of reducing distance from 10 to 6, which reduces the ability to correct errors - both potentially affecting the logical error rate  (LER) of this transversal CNOT operation  in opposite directions. 
    
\begin{figure}
\centering
\includegraphics[width=0.48\textwidth]{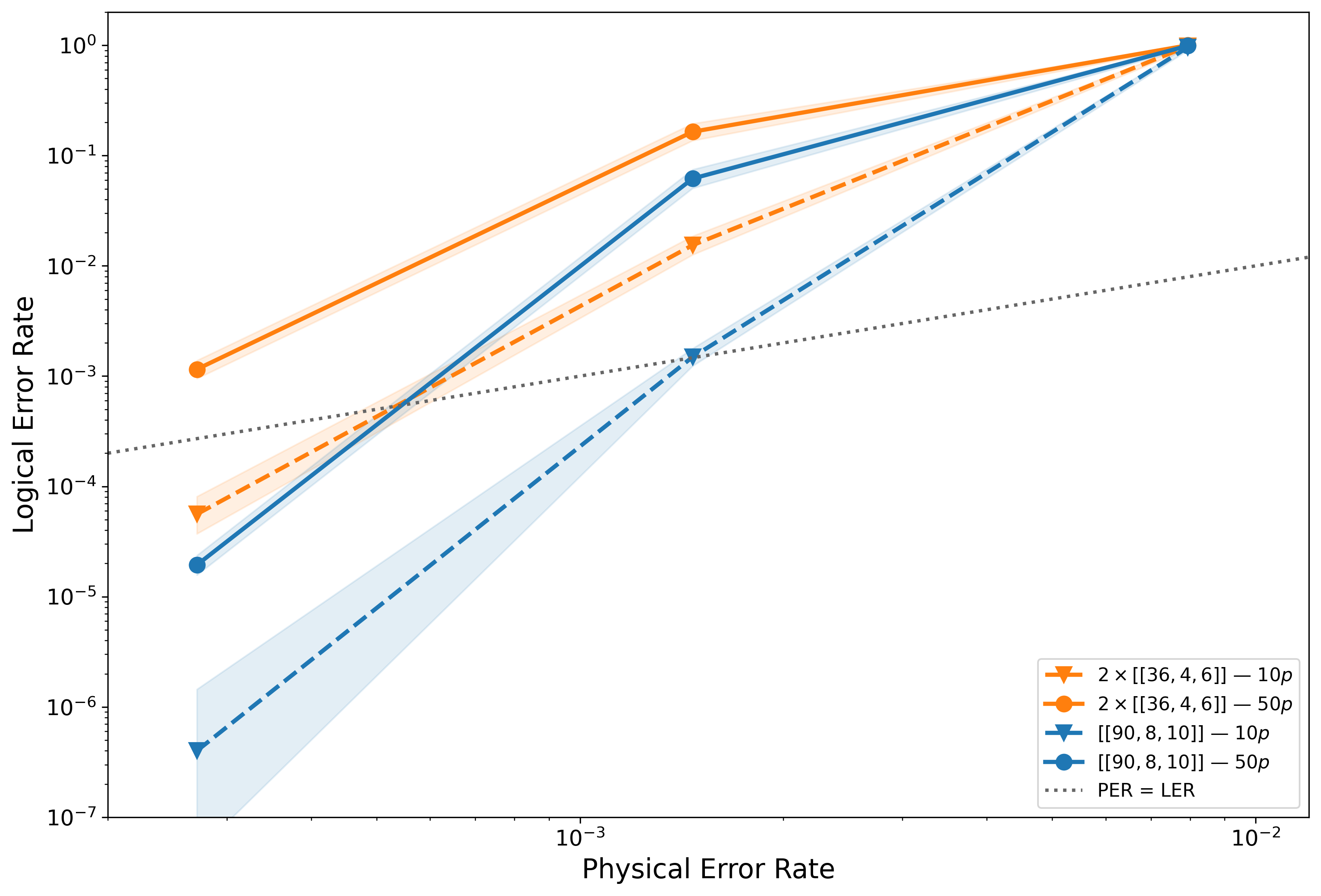}    
\scalebox{0.63}{
\begin{tabular}{cccccc}
\hline
Code & $p_{\rm ebit}$ & $\frac{kd^2}{n}$ & $p$ & LER & 95\% CI \\
\hline
$2\times[[36,4,6]]$ & $10p$ & $4.000$ & $2.71{\times}10^{-4}$
& $5.60{\times}10^{-5}$ & $[3.72{\times}10^{-5},8.09{\times}10^{-5}]$\\
& & & $1.468{\times}10^{-3}$
& $1.540{\times}10^{-2}$ & $[1.26{\times}10^{-2},1.86{\times}10^{-2}]$\\
& & & $7.943{\times}10^{-3}$
& $9.825{\times}10^{-1}$ & $[9.38{\times}10^{-1},9.98{\times}10^{-1}]$\\
\hline

$2\times[[36,4,6]]$ & $50p$ & $4.000$ & $2.71{\times}10^{-4}$
& $1.154{\times}10^{-3}$ & $[9.51{\times}10^{-4},1.39{\times}10^{-3}]$\\
& & & $1.468{\times}10^{-3}$
& $1.644{\times}10^{-1}$ & $[1.37{\times}10^{-1},1.95{\times}10^{-1}]$\\
& & & $7.943{\times}10^{-3}$
& $1.000$ & $[9.68{\times}10^{-1},1.000]$\\
\hline

$[[90,8,10]]$ & $10p$ & $8.889$ & $2.71{\times}10^{-4}$
& $4.00{\times}10^{-7}$ & $[4.84{\times}10^{-8},1.45{\times}10^{-6}]$\\
& & & $1.468{\times}10^{-3}$
& $1.491{\times}10^{-3}$ & $[1.24{\times}10^{-3},1.78{\times}10^{-3}]$\\
& & & $7.943{\times}10^{-3}$
& $9.615{\times}10^{-1}$ & $[9.04{\times}10^{-1},9.89{\times}10^{-1}]$\\
\hline

$[[90,8,10]]$ & $50p$ & $8.889$ & $2.71{\times}10^{-4}$
& $1.940{\times}10^{-5}$ & $[1.57{\times}10^{-5},2.37{\times}10^{-5}]$\\
& & & $1.468{\times}10^{-3}$
& $6.184{\times}10^{-2}$ & $[5.06{\times}10^{-2},7.47{\times}10^{-2}]$\\
& & & $7.943{\times}10^{-3}$
& $9.911{\times}10^{-1}$ & $[9.51{\times}10^{-1},1.00]$\\
\hline




\end{tabular}
}
  \caption{Comparison of the LERs (as the physical error rate $p$ varies) for the inter-node transversal CNOT operation using two different encodings for 8 logical qubits on each node: (i)  two $[[36,4,6]]$ blocks on each node, and (ii) one block of $[[90,8,10]]$ on each node. 
  }
\label{fig:cf-trans-cnot}
\end{figure}

    We study the  impact on the LER using a simulation based on  the Transversal Multiple Code Block Simulator (TMCBS)
library code from\cite{Stack2026Transversal} (developed for simulating non-local transversal CNOTs), which uses stim,\footnote{https://github.com/quantumlib/stim} and  we use sinter\footnote{https://pypi.org/project/sinter/}  for Monte Carlo sampling.  
Figure~\ref{fig:cf-trans-cnot}   shows that, when the probability of error in the ebit is $10\times$ the physical error rate (PER), i.e. $p_{ebit}=10p$, and even with much higher error rates in the ebits, namely, $p_{ebit}=50p$,   the LER is still much lower for the $[[90,8,10]]$ encoding than the 2x$[[36,4,6]]$ encoding, which means that the additional ebits did not increase the error rate sufficiently to offset the benefit of the larger encoding block. 
Despite  more  ebits, for the cases studied here, a higher efficiency  $kd^2/n$ yields a lower LER for the distributed transversal CNOT, e.g. an improvement by a factor of 140 for $p=2.71 \times 10^{-4}$ for $p_{ebit}=10p$.  
Note that the number of communication qubits required can be different from the number of ebits consumed (since we can initialise and reuse the communication qubits for newly formed ebits). For maximum parallelism, for the one $[[90,8,10]]$ block per node scheme, 90 communication qubits per node can support 90 concurrent ebits.
For the two $[[36,4,6]]$ blocks per node scheme, $\overline{CNOT}_{A',B'}$ and $\overline{CNOT}_{A'',B''}$ can be done in sequence (e.g., 36 ebits at a time requiring 36 communication qubits) or in parallel depending on the number of communication qubits available. 

\section{Encoding and Node Choices for GCZ} 
\label{sec:gcz}
 We consider the impact of different encodings on the LER of a larger circuit, namely, distributed versions of  global gates.
Global gates   are  efficient for some quantum architectures such as trapped ion qubits where entire arrays of qubits can  be targeted and pairwise qubit-qubit interactions   can be naturally realized~\cite{Wetering2021,maslovandnam2018}.
A global MS gate or operation (GMS) over a set of qubits $S$ is defined as follows: $GMS_S(\theta)=$
\[
 exp(-i\frac{\theta}{2} \sum_{i,j\in S, ~ i <j} X_i X_j)  = \prod_{i,j\in S, ~ i <j} exp(-i\frac{\theta}{2}X_i X_j)
 \]
where   $exp(-i\frac{\theta}{2}X_i X_j)$ is called the ``local'' MS gate acting on qubits $i$ and $j$, and can be viewed operationally as: 
$exp(-i\frac{\theta}{2}X_i X_j)$ = 
$(H_i \otimes H_j)  CNOT_{i\rightarrow j}  (I_i \otimes (R_Z(\theta))_j)  CNOT_{i\rightarrow j} (H_i \otimes H_j)$,
with  $R_Z(\theta) = \begin{bmatrix} e^{-i\frac{\theta}{2}} & 0 \\ 0 & e^{i\frac{\theta}{2}}  \end{bmatrix}$.
(Note that the local MS gate (or LMS gate, for short) is synmmetrical, i.e. $exp(-i\frac{\theta}{2}X_i X_j)$=$exp(-i\frac{\theta}{2}X_j X_i)$.) GCZ gates are equivalent (up to Clifford gates) to $GMS(\pi/2)$ gates~\cite{Wetering2021}.

\begin{figure*}[t]
    \centering
   \includegraphics[width=1.0\textwidth]{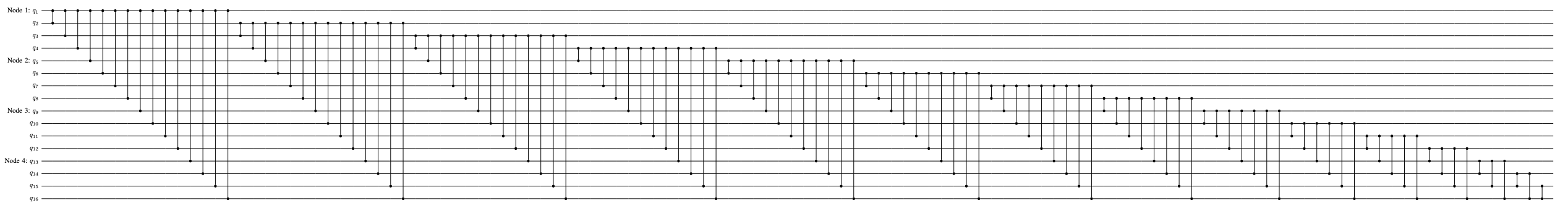}   
    \caption{Global gate operation (${\overline{GCZ}_L}_{16}$) over 4 nodes, with 120 ${\overline{CZ}_L}$s; $\mathrm{{\overline{GCZ}_L}}_{16} \;=\; \prod_{1 \le m < n \le 16} {\overline{CZ}_L}_{q_m,q_n}$}
    \label{fig:gcz120}
\end{figure*}

An illustration of a ${\overline{GCZ}_L}_{16}$ operation involving 16 (logical) qubits over four nodes, four (logical) qubits per node, is shown in Figure~\ref{fig:gcz120}, where there are 120 ${\overline{CZ}_L}$ operations, of which $16(16-1)/2 - 4*4(4-1)/2 = 96$ ${\overline{CZ}_L}$s are inter-node ${\overline{CZ}_L}$s.

If we focus on the first 54 (of the 120 ${\overline{CZ}_L}$s) shown in Figure~\ref{fig:gcz54}, 
note that we can reorder the operations and implement it using transversal CNOTs with an ancilla block, and fanning out from the ancilla blocks on each node to the computation qubits in the node, as shown in Appendix~A, Figure~\ref{fig:dgcz54} (where we set Node1 to be $C$, Node2 to be $T_1$, Node3 to be $T_2$ and Node4 to be $T_3$, with the ancilla blocks $A_1$, $A_2$ and $A_3$ on the Nodes 2, 3 and 4, respectively), so that all the 48 inter-node interactions among the 54 ${\overline{CZ}_L}$s can be realised using  three inter-node transversal CNOT operations as shown in Figure~\ref{fig:three-cnot-blocks}.\footnote{The operations can be done using distributed fanouts as shown in~\cite{transversalfanout}, but in this paper, we use only transversal non-local CNOTs followed by local fanouts.}  (For simplicity, we mostly ignore the  details of the uncomputation (with LER  $l_{unc}$) of the ancilla blocks in the discussion below.)

\begin{figure}[t]
    \centering
    \resizebox{\linewidth}{!}{%
\begin{quantikz}[row sep=0.22cm, column sep=0.10cm]
\lstick{\text{Node 1: }$q_1$} & \ctrl{1} & \ctrl{2} & \ctrl{3} & \ctrl{4} & \ctrl{5} & \ctrl{6} & \ctrl{7} & \ctrl{8} & \ctrl{9} & \ctrl{10} & \ctrl{11} & \ctrl{12} & \ctrl{13} & \ctrl{14} & \ctrl{15} & \qw & \qw & \qw & \qw & \qw & \qw & \qw & \qw & \qw & \qw & \qw & \qw & \qw & \qw & \qw & \qw & \qw & \qw & \qw & \qw & \qw & \qw & \qw & \qw & \qw & \qw & \qw & \qw & \qw & \qw & \qw & \qw & \qw & \qw & \qw & \qw & \qw & \qw & \qw & \qw \\
\lstick{$q_{2}$} & \control{} & \qw & \qw & \qw & \qw & \qw & \qw & \qw & \qw & \qw & \qw & \qw & \qw & \qw & \qw & \ctrl{1} & \ctrl{2} & \ctrl{3} & \ctrl{4} & \ctrl{5} & \ctrl{6} & \ctrl{7} & \ctrl{8} & \ctrl{9} & \ctrl{10} & \ctrl{11} & \ctrl{12} & \ctrl{13} & \ctrl{14} & \qw & \qw & \qw & \qw & \qw & \qw & \qw & \qw & \qw & \qw & \qw & \qw & \qw & \qw & \qw & \qw & \qw & \qw & \qw & \qw & \qw & \qw & \qw & \qw & \qw & \qw \\
\lstick{$q_{3}$} & \qw & \control{} & \qw & \qw & \qw & \qw & \qw & \qw & \qw & \qw & \qw & \qw & \qw & \qw & \qw & \control{} & \qw & \qw & \qw & \qw & \qw & \qw & \qw & \qw & \qw & \qw & \qw & \qw & \qw & \ctrl{1} & \ctrl{2} & \ctrl{3} & \ctrl{4} & \ctrl{5} & \ctrl{6} & \ctrl{7} & \ctrl{8} & \ctrl{9} & \ctrl{10} & \ctrl{11} & \ctrl{12} & \ctrl{13} & \qw & \qw & \qw & \qw & \qw & \qw & \qw & \qw & \qw & \qw & \qw & \qw & \qw \\
\lstick{$q_{4}$} & \qw & \qw & \control{} & \qw & \qw & \qw & \qw & \qw & \qw & \qw & \qw & \qw & \qw & \qw & \qw & \qw & \control{} & \qw & \qw & \qw & \qw & \qw & \qw & \qw & \qw & \qw & \qw & \qw & \qw & \control{} & \qw & \qw & \qw & \qw & \qw & \qw & \qw & \qw & \qw & \qw & \qw & \qw & \ctrl{1} & \ctrl{2} & \ctrl{3} & \ctrl{4} & \ctrl{5} & \ctrl{6} & \ctrl{7} & \ctrl{8} & \ctrl{9} & \ctrl{10} & \ctrl{11} & \ctrl{12} & \qw \\
\lstick{\text{Node 2: }$q_5$} & \qw & \qw & \qw & \control{} & \qw & \qw & \qw & \qw & \qw & \qw & \qw & \qw & \qw & \qw & \qw & \qw & \qw & \control{} & \qw & \qw & \qw & \qw & \qw & \qw & \qw & \qw & \qw & \qw & \qw & \qw & \control{} & \qw & \qw & \qw & \qw & \qw & \qw & \qw & \qw & \qw & \qw & \qw & \control{} & \qw & \qw & \qw & \qw & \qw & \qw & \qw & \qw & \qw & \qw & \qw & \qw \\
\lstick{$q_{6}$} & \qw & \qw & \qw & \qw & \control{} & \qw & \qw & \qw & \qw & \qw & \qw & \qw & \qw & \qw & \qw & \qw & \qw & \qw & \control{} & \qw & \qw & \qw & \qw & \qw & \qw & \qw & \qw & \qw & \qw & \qw & \qw & \control{} & \qw & \qw & \qw & \qw & \qw & \qw & \qw & \qw & \qw & \qw & \qw & \control{} & \qw & \qw & \qw & \qw & \qw & \qw & \qw & \qw & \qw & \qw & \qw \\
\lstick{$q_{7}$} & \qw & \qw & \qw & \qw & \qw & \control{} & \qw & \qw & \qw & \qw & \qw & \qw & \qw & \qw & \qw & \qw & \qw & \qw & \qw & \control{} & \qw & \qw & \qw & \qw & \qw & \qw & \qw & \qw & \qw & \qw & \qw & \qw & \control{} & \qw & \qw & \qw & \qw & \qw & \qw & \qw & \qw & \qw & \qw & \qw & \control{} & \qw & \qw & \qw & \qw & \qw & \qw & \qw & \qw & \qw & \qw \\
\lstick{$q_{8}$} & \qw & \qw & \qw & \qw & \qw & \qw & \control{} & \qw & \qw & \qw & \qw & \qw & \qw & \qw & \qw & \qw & \qw & \qw & \qw & \qw & \control{} & \qw & \qw & \qw & \qw & \qw & \qw & \qw & \qw & \qw & \qw & \qw & \qw & \control{} & \qw & \qw & \qw & \qw & \qw & \qw & \qw & \qw & \qw & \qw & \qw & \control{} & \qw & \qw & \qw & \qw & \qw & \qw & \qw & \qw & \qw \\
\lstick{\text{Node 3: }$q_9$} & \qw & \qw & \qw & \qw & \qw & \qw & \qw & \control{} & \qw & \qw & \qw & \qw & \qw & \qw & \qw & \qw & \qw & \qw & \qw & \qw & \qw & \control{} & \qw & \qw & \qw & \qw & \qw & \qw & \qw & \qw & \qw & \qw & \qw & \qw & \control{} & \qw & \qw & \qw & \qw & \qw & \qw & \qw & \qw & \qw & \qw & \qw & \control{} & \qw & \qw & \qw & \qw & \qw & \qw & \qw & \qw \\
\lstick{$q_{10}$} & \qw & \qw & \qw & \qw & \qw & \qw & \qw & \qw & \control{} & \qw & \qw & \qw & \qw & \qw & \qw & \qw & \qw & \qw & \qw & \qw & \qw & \qw & \control{} & \qw & \qw & \qw & \qw & \qw & \qw & \qw & \qw & \qw & \qw & \qw & \qw & \control{} & \qw & \qw & \qw & \qw & \qw & \qw & \qw & \qw & \qw & \qw & \qw & \control{} & \qw & \qw & \qw & \qw & \qw & \qw & \qw \\
\lstick{$q_{11}$} & \qw & \qw & \qw & \qw & \qw & \qw & \qw & \qw & \qw & \control{} & \qw & \qw & \qw & \qw & \qw & \qw & \qw & \qw & \qw & \qw & \qw & \qw & \qw & \control{} & \qw & \qw & \qw & \qw & \qw & \qw & \qw & \qw & \qw & \qw & \qw & \qw & \control{} & \qw & \qw & \qw & \qw & \qw & \qw & \qw & \qw & \qw & \qw & \qw & \control{} & \qw & \qw & \qw & \qw & \qw & \qw \\
\lstick{$q_{12}$} & \qw & \qw & \qw & \qw & \qw & \qw & \qw & \qw & \qw & \qw & \control{} & \qw & \qw & \qw & \qw & \qw & \qw & \qw & \qw & \qw & \qw & \qw & \qw & \qw & \control{} & \qw & \qw & \qw & \qw & \qw & \qw & \qw & \qw & \qw & \qw & \qw & \qw & \control{} & \qw & \qw & \qw & \qw & \qw & \qw & \qw & \qw & \qw & \qw & \qw & \control{} & \qw & \qw & \qw & \qw & \qw \\
\lstick{\text{Node 4: }$q_{13}$} & \qw & \qw & \qw & \qw & \qw & \qw & \qw & \qw & \qw & \qw & \qw & \control{} & \qw & \qw & \qw & \qw & \qw & \qw & \qw & \qw & \qw & \qw & \qw & \qw & \qw & \control{} & \qw & \qw & \qw & \qw & \qw & \qw & \qw & \qw & \qw & \qw & \qw & \qw & \control{} & \qw & \qw & \qw & \qw & \qw & \qw & \qw & \qw & \qw & \qw & \qw & \control{} & \qw & \qw & \qw & \qw \\
\lstick{$q_{14}$} & \qw & \qw & \qw & \qw & \qw & \qw & \qw & \qw & \qw & \qw & \qw & \qw & \control{} & \qw & \qw & \qw & \qw & \qw & \qw & \qw & \qw & \qw & \qw & \qw & \qw & \qw & \control{} & \qw & \qw & \qw & \qw & \qw & \qw & \qw & \qw & \qw & \qw & \qw & \qw & \control{} & \qw & \qw & \qw & \qw & \qw & \qw & \qw & \qw & \qw & \qw & \qw & \control{} & \qw & \qw & \qw \\
\lstick{$q_{15}$} & \qw & \qw & \qw & \qw & \qw & \qw & \qw & \qw & \qw & \qw & \qw & \qw & \qw & \control{} & \qw & \qw & \qw & \qw & \qw & \qw & \qw & \qw & \qw & \qw & \qw & \qw & \qw & \control{} & \qw & \qw & \qw & \qw & \qw & \qw & \qw & \qw & \qw & \qw & \qw & \qw & \control{} & \qw & \qw & \qw & \qw & \qw & \qw & \qw & \qw & \qw & \qw & \qw & \control{} & \qw & \qw \\
\lstick{$q_{16}$} & \qw & \qw & \qw & \qw & \qw & \qw & \qw & \qw & \qw & \qw & \qw & \qw & \qw & \qw & \control{} & \qw & \qw & \qw & \qw & \qw & \qw & \qw & \qw & \qw & \qw & \qw & \qw & \qw & \control{} & \qw & \qw & \qw & \qw & \qw & \qw & \qw & \qw & \qw & \qw & \qw & \qw & \control{} & \qw & \qw & \qw & \qw & \qw & \qw & \qw & \qw & \qw & \qw & \qw & \control{} & \qw \\
\end{quantikz}
}
    \caption{The first 54 ${\overline{CZ}_L}$s, from Figure~\ref{fig:gcz120}.}
    \label{fig:gcz54}
\end{figure}
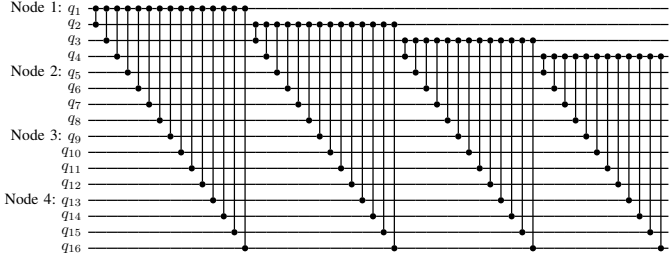

\begin{figure}[t]
\centering
\scalebox{0.65}{
\begin{quantikz}
\lstick{$C:\ q_1$} & \ctrl{4} \gategroup[wires=8,steps=4,style={dashed,red}, label style={label position=above}]{trans. CNOT1} & \qw & \qw & \qw & \ctrl{8}  \gategroup[wires=12,steps=4,style={dashed,red}, label style={label position=above}]{trans. CNOT2} & \qw & \qw & \qw & \ctrl{12} \gategroup[wires=16,steps=4,style={dashed,red}, label style={label position=above}]{trans. CNOT3} & \qw & \qw & \qw & \qw \\
\lstick{$q_2$} & \qw & \ctrl{4} & \qw & \qw & \qw & \ctrl{8} & \qw & \qw & \qw & \ctrl{12} & \qw & \qw & \qw \\
\lstick{$q_3$} & \qw & \qw & \ctrl{4} & \qw & \qw & \qw & \ctrl{8} & \qw & \qw & \qw & \ctrl{12} & \qw & \qw \\
\lstick{$q_4$} & \qw & \qw & \qw & \ctrl{4} & \qw & \qw & \qw & \ctrl{8} & \qw & \qw & \qw & \ctrl{12} & \qw \\
\lstick{$A_1:\ a_{T_1,1}$} & \targ{} & \qw & \qw & \qw & \qw & \qw & \qw & \qw & \qw & \qw & \qw & \qw & \qw \\
\lstick{$a_{T_1,2}$} & \qw & \targ{} & \qw & \qw & \qw & \qw & \qw & \qw & \qw & \qw & \qw & \qw & \qw \\
\lstick{$a_{T_1,3}$} & \qw & \qw & \targ{} & \qw & \qw & \qw & \qw & \qw & \qw & \qw & \qw & \qw & \qw \\
\lstick{$a_{T_1,4}$} & \qw & \qw & \qw & \targ{} & \qw & \qw & \qw & \qw & \qw & \qw & \qw & \qw & \qw \\
\lstick{$A_2:\ a_{T_2,1}$} & \qw & \qw & \qw & \qw & \targ{} & \qw & \qw & \qw & \qw & \qw & \qw & \qw & \qw \\
\lstick{$a_{T_2,2}$} & \qw & \qw & \qw & \qw & \qw & \targ{} & \qw & \qw & \qw & \qw & \qw & \qw & \qw \\
\lstick{$a_{T_2,3}$} & \qw & \qw & \qw & \qw & \qw & \qw & \targ{} & \qw & \qw & \qw & \qw & \qw & \qw \\
\lstick{$a_{T_2,4}$} & \qw & \qw & \qw & \qw & \qw & \qw & \qw & \targ{} & \qw & \qw & \qw & \qw & \qw \\
\lstick{$A_3:\ a_{T_3,1}$} & \qw & \qw & \qw & \qw & \qw & \qw & \qw & \qw & \targ{} & \qw & \qw & \qw & \qw \\
\lstick{$a_{T_3,2}$} & \qw & \qw & \qw & \qw & \qw & \qw & \qw & \qw & \qw & \targ{} & \qw & \qw & \qw \\
\lstick{$a_{T_3,3}$} & \qw & \qw & \qw & \qw & \qw & \qw & \qw & \qw & \qw & \qw & \targ{} & \qw & \qw \\
\lstick{$a_{T_3,4}$} & \qw & \qw & \qw & \qw & \qw & \qw & \qw & \qw & \qw & \qw & \qw & \targ{} & \qw 
\end{quantikz}
}
\caption{Three sets of logical CNOT operations between the computational qubits and ancilla qubits; logical-level representation indicates   corresponding physical transversal operation on the encoded blocks.}
\label{fig:three-cnot-blocks}
\end{figure}
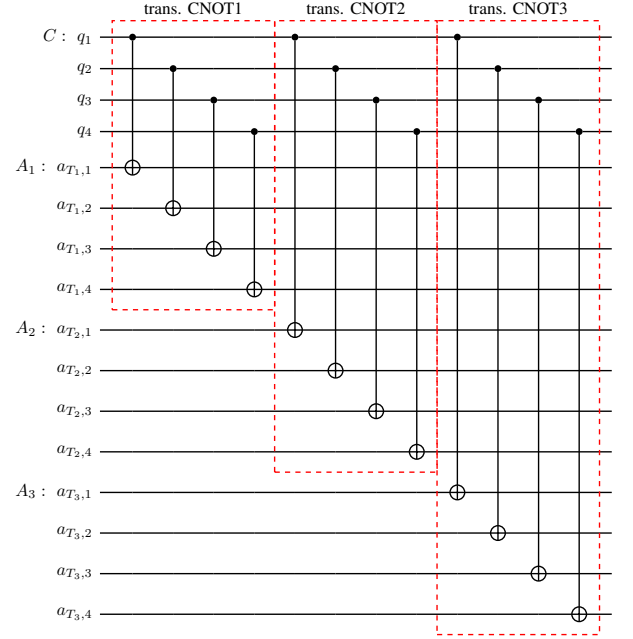

Similarly, for the other inter-node transversal ${\overline{CZ}_L}$s from Figure 4, another three inter-node  transversal CNOTs are needed: one between   Node2 and  Node3, and another between Node2 and Node4, and one more  inter-node transversal CNOT between Node3 and Node4. Hence, using this scheme, six inter-node transversal CNOTs are required for ${\overline{GCZ}_L}_{16}$ over the four nodes, if we leave out uncomputation.
For each set of  four logical qubits on a node, BB code options include   the $[[36,4,6]]$ or $[[54,4,8]]$ encoding. 
Hence, using $[[54,4,8]]$ encoding,
an approximate estimate of the LER of the distributed ${\overline{GCZ}_L}_{16}$,  with 16 logical qubits, over 4 nodes, 4 logical qubits per node, i.e., configuration $(4+4+4+4)_{[[54,4,8]]}$, denoted by $LER({\overline{GCZ}_L}_{16},(4+4+4+4)_{[[54,4,8]]})$, is as follows:\footnote{
Assuming independent operation-level logical failures, viewing $LER$ as probability $p$ of a logical error, we have:
\begin{align*}
& p_{[{\overline{GCZ}_L}_{16},(4+4+4+4)_{[[54,4,8]]} ~w/o~ unc]} \\
& =1-[(1-p_{tCNOT})^6(1-p_{localOps})]  \approx 6p_{tCNOT} + p_{localOps}
\end{align*}
where the last line approximation is for small  enough LER or probabilities $p_{tCNOT}$ and $p_{localOps}$.
}
\begin{align*}
& LER({\overline{GCZ}_L}_{16},(4+4+4+4)_{[[54,4,8]]})  \\
& \approx  6LER(tCNOT_{[[54,4,8]]}) + LER(localOps_{[[54,4,8]]}) + l_{unc}
\end{align*}
where $LER(tCNOT_{[[54,4,8]]})$ denotes the LER of the inter-node transversal CNOT between two nodes, each of 4 logical qubits encoded using  $[[54,4,8]]$, and $LER(localOps_{[[54,4,8]]})$ denotes the LER of the rest of this circuit, namely, the local operations. Similarly, using the $[[36,4,6]]$ encoding, we have 

\begin{align*}
& LER({\overline{GCZ}_L}_{16},(4+4+4+4)_{[[36,4,6]]})   \\
& \approx 6LER(tCNOT_{[[36,4,6]]}) + LER(localOps_{[[36,4,6]]}) + l_{unc}
\end{align*}

\begin{figure}[ht!]
\centering
  \includegraphics[width=0.48\textwidth]{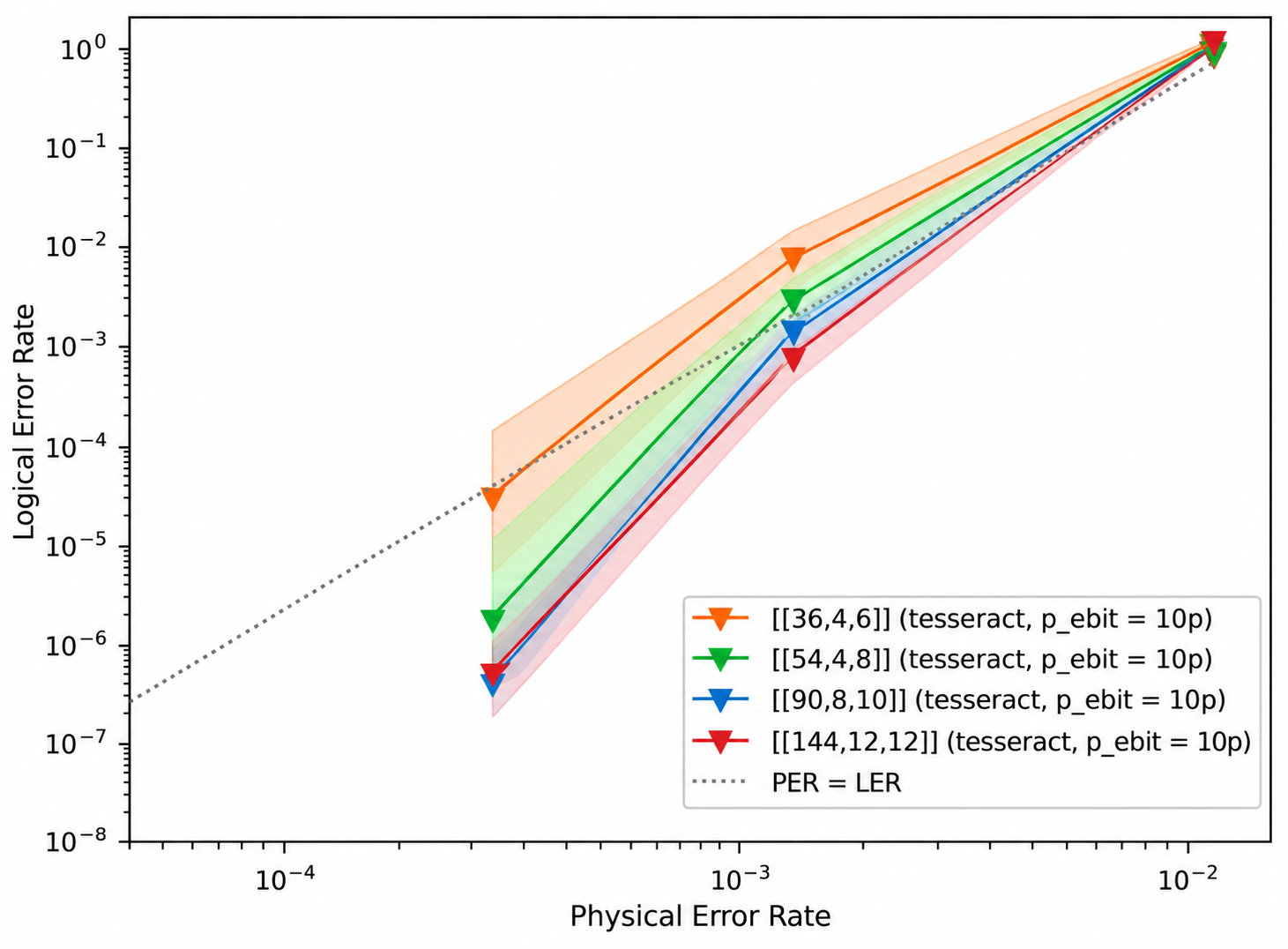}
  \centering
\scalebox{0.68}{
\begin{tabular}{c|c|c|c}
\hline
Code & $p$ & Logical Error Rate & 95\% CI \\
\hline

$[[144,12,12]]$
& $2.710\times10^{-4}$
& $4.000\times10^{-7}$
& $[4.844\times10^{-8},\,1.445\times10^{-6}]$ \\


$[[90,8,10]]$
& $2.710\times10^{-4}$
& $4.000\times10^{-7}$
& $[4.844\times10^{-8},\,1.445\times10^{-6}]$ \\

$[[54,4,8]]$
& $2.710\times10^{-4}$
& $1.400\times10^{-6}$
& $[5.629\times10^{-7},\,2.885\times10^{-6}]$ \\

$[[36,4,6]]$
& $2.710\times10^{-4}$
& $2.587\times10^{-5}$
& $[2.157\times10^{-5},\,3.078\times10^{-5}]$ \\

\hline

$[[144,12,12]]$
& $1.468\times10^{-3}$
& $8.277\times10^{-4}$
& $[6.821\times10^{-4},\,9.950\times10^{-4}]$ \\


$[[90,8,10]]$
& $1.468\times10^{-3}$
& $1.491\times10^{-3}$
& $[1.238\times10^{-3},\,1.782\times10^{-3}]$ \\

$[[54,4,8]]$
& $1.468\times10^{-3}$
& $2.561\times10^{-3}$
& $[2.090\times10^{-3},\,3.105\times10^{-3}]$ \\

$[[36,4,6]]$
& $1.468\times10^{-3}$
& $6.249\times10^{-3}$
& $[5.134\times10^{-3},\,7.534\times10^{-3}]$ \\

\hline

$[[144,12,12]]$
& $7.943\times10^{-3}$
& $1.000\times10^{0}$
& $[9.638\times10^{-1},\,1]$ \\


$[[90,8,10]]$
& $7.943\times10^{-3}$
& $9.615\times10^{-1}$
& $[9.044\times10^{-1},\,9.894\times10^{-1}]$ \\

$[[54,4,8]]$
& $7.943\times10^{-3}$
& $8.130\times10^{-1}$
& $[7.328\times10^{-1},\,8.776\times10^{-1}]$ \\

$[[36,4,6]]$
& $7.943\times10^{-3}$
& $8.487\times10^{-1}$
& $[7.715\times10^{-1},\,9.078\times10^{-1}]$ \\

\hline
\end{tabular}
}
  \caption{Comparison of the LERs of inter-node transversal CNOTs involving two blocks of  $[[144,12,12]]$, 
  two blocks of  $[[90,8,10]]$, two blocks of $[[54,4,8]]$ and  two blocks of $[[36,4,6]]$, over two nodes, one block per node, with ebit error rate $p_{ebit}=10p$. The table shows the values plotted with   95\% confidence intervals.}
\label{fig:cf-trans-cnot-diffcap}
\end{figure}

From Figure~\ref{fig:cf-trans-cnot-diffcap}, using $[[54,4,8]]$ blocks, one per node, an inter-node transversal CNOT can have a lower LER than $[[36,4,6]]$ blocks, despite using more ebits, i.e. $LER(tCNOT_{[[54,4,8]]}) \leq LER(tCNOT_{[[36,4,6]]})$; thus:
\begin{align*}
& LER({\overline{GCZ}_L}_{16},(4+4+4+4)_{[[54,4,8]]}) \\
& \leq LER({\overline{GCZ}_L}_{16},(4+4+4+4)_{[[36,4,6]]}) 
\end{align*}
despite the higher number of physical ebits required (i.e., 6x18=108 ebits more, and with $p_{ebit}=10p$), and assuming $LER(localOps_{[[54,4,8]]})$ close to $LER(localOps_{[[36,4,6]]})$, and perhaps the local operations are even better with $[[54,4,8]]$ blocks due to the larger distance. Due to symmetry, with uncomputation, the inequality could still hold.\footnote{With uncomputation, for example, we would have:  $LER({\overline{GCZ}_L}_{16},(4+4+4+4)_{[[54,4,8]]})   \approx  12LER(tCNOT_{[[54,4,8]]}) + LER(localOps_{[[54,4,8]]})$. Similarly, with $LER({\overline{GCZ}_L}_{16},(4+4+4+4)_{[[36,4,6]]})$.}

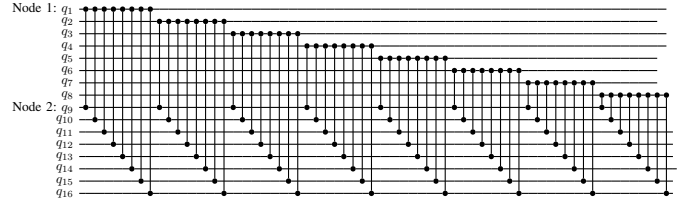
\begin{figure}[t]
    \centering
    \resizebox{1.0\linewidth}{!}{%
\begin{quantikz}[row sep=0.18cm,column sep=0.10cm]
\lstick{\text{Node 1: }$q_1$} & \ctrl{8} & \ctrl{9} & \ctrl{10} & \ctrl{11} & \ctrl{12} & \ctrl{13} & \ctrl{14} & \ctrl{15} & \qw & \qw & \qw & \qw & \qw & \qw & \qw & \qw & \qw & \qw & \qw & \qw & \qw & \qw & \qw & \qw & \qw & \qw & \qw & \qw & \qw & \qw & \qw & \qw & \qw & \qw & \qw & \qw & \qw & \qw & \qw & \qw & \qw & \qw & \qw & \qw & \qw & \qw & \qw & \qw & \qw & \qw & \qw & \qw & \qw & \qw & \qw & \qw & \qw & \qw & \qw & \qw & \qw & \qw & \qw & \qw \\
\lstick{$q_2$} & \qw & \qw & \qw & \qw & \qw & \qw & \qw & \qw & \ctrl{7} & \ctrl{8} & \ctrl{9} & \ctrl{10} & \ctrl{11} & \ctrl{12} & \ctrl{13} & \ctrl{14} & \qw & \qw & \qw & \qw & \qw & \qw & \qw & \qw & \qw & \qw & \qw & \qw & \qw & \qw & \qw & \qw & \qw & \qw & \qw & \qw & \qw & \qw & \qw & \qw & \qw & \qw & \qw & \qw & \qw & \qw & \qw & \qw & \qw & \qw & \qw & \qw & \qw & \qw & \qw & \qw & \qw & \qw & \qw & \qw & \qw & \qw & \qw \\
\lstick{$q_3$} & \qw & \qw & \qw & \qw & \qw & \qw & \qw & \qw & \qw & \qw & \qw & \qw & \qw & \qw & \qw & \qw & \ctrl{6} & \ctrl{7} & \ctrl{8} & \ctrl{9} & \ctrl{10} & \ctrl{11} & \ctrl{12} & \ctrl{13} & \qw & \qw & \qw & \qw & \qw & \qw & \qw & \qw & \qw & \qw & \qw & \qw & \qw & \qw & \qw & \qw & \qw & \qw & \qw & \qw & \qw & \qw & \qw & \qw & \qw & \qw & \qw & \qw & \qw & \qw & \qw & \qw & \qw & \qw & \qw & \qw & \qw & \qw & \qw & \qw \\
\lstick{$q_4$} & \qw & \qw & \qw & \qw & \qw & \qw & \qw & \qw & \qw & \qw & \qw & \qw & \qw & \qw & \qw & \qw & \qw & \qw & \qw & \qw & \qw & \qw & \qw & \qw & \ctrl{5} & \ctrl{6} & \ctrl{7} & \ctrl{8} & \ctrl{9} & \ctrl{10} & \ctrl{11} & \ctrl{12} & \qw & \qw & \qw & \qw & \qw & \qw & \qw & \qw & \qw & \qw & \qw & \qw & \qw & \qw & \qw & \qw & \qw & \qw & \qw & \qw & \qw & \qw & \qw & \qw & \qw & \qw & \qw & \qw & \qw & \qw & \qw & \qw \\
\lstick{$q_5$} & \qw & \qw & \qw & \qw & \qw & \qw & \qw & \qw & \qw & \qw & \qw & \qw & \qw & \qw & \qw & \qw & \qw & \qw & \qw & \qw & \qw & \qw & \qw & \qw & \qw & \qw & \qw & \qw & \qw & \qw & \qw & \qw & \ctrl{4} & \ctrl{5} & \ctrl{6} & \ctrl{7} & \ctrl{8} & \ctrl{9} & \ctrl{10} & \ctrl{11} & \qw & \qw & \qw & \qw & \qw & \qw & \qw & \qw & \qw & \qw & \qw & \qw & \qw & \qw & \qw & \qw & \qw & \qw & \qw & \qw & \qw & \qw & \qw \\
\lstick{$q_6$} & \qw & \qw & \qw & \qw & \qw & \qw & \qw & \qw & \qw & \qw & \qw & \qw & \qw & \qw & \qw & \qw & \qw & \qw & \qw & \qw & \qw & \qw & \qw & \qw & \qw & \qw & \qw & \qw & \qw & \qw & \qw & \qw & \qw & \qw & \qw & \qw & \qw & \qw & \qw & \qw & \ctrl{3} & \ctrl{4} & \ctrl{5} & \ctrl{6} & \ctrl{7} & \ctrl{8} & \ctrl{9} & \ctrl{10} & \qw & \qw & \qw & \qw & \qw & \qw & \qw & \qw & \qw & \qw & \qw & \qw & \qw & \qw & \qw \\
\lstick{$q_7$} & \qw & \qw & \qw & \qw & \qw & \qw & \qw & \qw & \qw & \qw & \qw & \qw & \qw & \qw & \qw & \qw & \qw & \qw & \qw & \qw & \qw & \qw & \qw & \qw & \qw & \qw & \qw & \qw & \qw & \qw & \qw & \qw & \qw & \qw & \qw & \qw & \qw & \qw & \qw & \qw & \qw & \qw & \qw & \qw & \qw & \qw & \qw &  \qw & \ctrl{2} & \ctrl{3} & \ctrl{4} & \ctrl{5} & \ctrl{6} & \ctrl{7} & \ctrl{8} & \ctrl{9} & \qw & \qw & \qw & \qw & \qw & \qw & \qw \\
\lstick{$q_8$} & \qw & \qw & \qw & \qw & \qw & \qw & \qw & \qw & \qw & \qw & \qw & \qw & \qw & \qw & \qw & \qw & \qw & \qw & \qw & \qw & \qw & \qw & \qw & \qw & \qw & \qw & \qw & \qw & \qw & \qw & \qw & \qw & \qw & \qw & \qw & \qw & \qw & \qw & \qw & \qw & \qw & \qw & \qw & \qw & \qw & \qw & \qw & \qw & \qw & \qw & \qw & \qw & \qw & \qw & \qw & \qw & \ctrl{1} & \ctrl{2} & \ctrl{3} & \ctrl{4} & \ctrl{5} & \ctrl{6} & \ctrl{7} & \ctrl{8} \\
\lstick{\text{Node 2: }$q_9$} & \control{} & \qw & \qw & \qw & \qw & \qw & \qw & \qw & \control{} & \qw & \qw & \qw & \qw & \qw & \qw & \qw & \control{} & \qw & \qw & \qw & \qw & \qw & \qw & \qw & \control{} & \qw & \qw & \qw & \qw & \qw & \qw & \qw & \control{} & \qw & \qw & \qw & \qw & \qw & \qw & \qw & \control{} & \qw & \qw & \qw & \qw & \qw & \qw & \qw & \control{} & \qw & \qw & \qw & \qw & \qw & \qw & \qw & \control{} & \qw & \qw & \qw & \qw & \qw & \qw & \qw \\
\lstick{$q_{10}$} & \qw & \control{} & \qw & \qw & \qw & \qw & \qw & \qw & \qw & \control{} & \qw & \qw & \qw & \qw & \qw & \qw & \qw & \control{} & \qw & \qw & \qw & \qw & \qw & \qw & \qw & \control{} & \qw & \qw & \qw & \qw & \qw & \qw & \qw & \control{} & \qw & \qw & \qw & \qw & \qw & \qw & \qw & \control{} & \qw & \qw & \qw & \qw & \qw & \qw & \qw & \control{} & \qw & \qw & \qw & \qw & \qw & \qw & \qw & \control{} & \qw & \qw & \qw & \qw & \qw & \qw \\
\lstick{$q_{11}$} & \qw & \qw & \control{} & \qw & \qw & \qw & \qw & \qw & \qw & \qw & \control{} & \qw & \qw & \qw & \qw & \qw & \qw & \qw & \control{} & \qw & \qw & \qw & \qw & \qw & \qw & \qw & \control{} & \qw & \qw & \qw & \qw & \qw & \qw & \qw & \control{} & \qw & \qw & \qw & \qw & \qw & \qw & \qw & \control{} & \qw & \qw & \qw & \qw & \qw & \qw & \qw & \control{} & \qw & \qw & \qw & \qw & \qw & \qw & \qw & \control{} & \qw & \qw & \qw & \qw & \qw & \qw \\
\lstick{$q_{12}$} & \qw & \qw & \qw & \control{} & \qw & \qw & \qw & \qw & \qw & \qw & \qw & \control{} & \qw & \qw & \qw & \qw & \qw & \qw & \qw & \control{} & \qw & \qw & \qw & \qw & \qw & \qw & \qw & \control{} & \qw & \qw & \qw & \qw & \qw & \qw & \qw & \control{} & \qw & \qw & \qw & \qw & \qw & \qw & \qw & \control{} & \qw & \qw & \qw & \qw & \qw & \qw & \qw & \control{} & \qw & \qw & \qw & \qw & \qw & \qw & \qw & \control{} & \qw & \qw & \qw & \qw & \qw   \\
\lstick{$q_{13}$} & \qw & \qw & \qw & \qw & \control{} & \qw & \qw & \qw & \qw & \qw & \qw & \qw & \control{} & \qw & \qw & \qw & \qw & \qw & \qw & \qw  & \control{} & \qw & \qw & \qw & \qw & \qw & \qw & \qw & \control{} & \qw & \qw & \qw & \qw & \qw & \qw & \qw & \control{} & \qw & \qw & \qw & \qw & \qw & \qw & \qw & \control{} & \qw & \qw & \qw & \qw & \qw & \qw & \qw & \control{} & \qw & \qw & \qw & \qw & \qw & \qw & \qw & \control{} & \qw & \qw & \qw & \qw  \\
\lstick{$q_{14}$} & \qw & \qw & \qw & \qw & \qw & \control{} & \qw & \qw & \qw & \qw & \qw & \qw & \qw & \control{} & \qw & \qw & \qw & \qw & \qw & \qw & \qw & \control{} & \qw & \qw & \qw & \qw & \qw & \qw & \qw & \control{} & \qw & \qw & \qw & \qw & \qw & \qw & \qw & \control{} & \qw & \qw & \qw & \qw & \qw & \qw & \qw & \control{} & \qw & \qw & \qw & \qw & \qw & \qw & \qw & \control{} & \qw & \qw & \qw & \qw & \qw & \qw & \qw & \control{} & \qw & \qw & \qw & \qw   \\
\lstick{$q_{15}$} & \qw & \qw & \qw & \qw & \qw & \qw & \control{} & \qw & \qw & \qw & \qw & \qw & \qw & \qw & \control{} & \qw & \qw & \qw & \qw & \qw & \qw & \qw & \control{} & \qw & \qw & \qw & \qw & \qw & \qw & \qw & \control{} & \qw & \qw & \qw & \qw & \qw & \qw & \qw & \control{} & \qw & \qw & \qw & \qw & \qw & \qw & \qw & \control{} & \qw & \qw & \qw & \qw & \qw & \qw & \qw & \control{} & \qw & \qw & \qw & \qw & \qw & \qw & \qw & \control{} & \qw & \qw \\
\lstick{$q_{16}$} & \qw & \qw & \qw & \qw & \qw & \qw & \qw & \control{} & \qw & \qw & \qw & \qw & \qw & \qw & \qw & \control{} & \qw & \qw & \qw & \qw & \qw & \qw & \qw & \control{} & \qw & \qw & \qw & \qw & \qw & \qw & \qw & \control{} & \qw & \qw & \qw & \qw & \qw & \qw & \qw & \control{} & \qw & \qw & \qw & \qw & \qw & \qw & \qw & \control{} & \qw & \qw & \qw & \qw & \qw & \qw & \qw & \control{} & \qw & \qw & \qw & \qw & \qw & \qw & \qw & \control{} & \qw
\end{quantikz}
    }
    \caption{Showing only the 64 inter-node ${\overline{CZ}_L}$s for ${\overline{GCZ}_L}_{16}$ over two nodes. }
    \label{fig:non-local-part-gcz-2nodes}
\end{figure}

Over two nodes, the global gate operation has inter-node operations shown in Figure~\ref{fig:non-local-part-gcz-2nodes}. With a similar reordering as above, each node can use blocks for 8 logical qubits, requiring one inter-node transversal CNOT, e.g. $[[90,8,10]]$, followed by local fanouts from ancilla qubits to the computation qubits.\footnote{One can also consider an approach with the communication qubits effectively  acting as the ancilla qubits.}

Similar arguments can be extended to say the ${\overline{GCZ}_L}_{24}$ case over two nodes, say, with 12 logical qubits per node. One might use $[[144,12,12]]$ blocks rather than 3x$[[36,4,6]]$ blocks despite the 144-3x36=36 more physical ebits required.

We can also consider choices for the transversal operation with nodes of different capacities. For example, suppose for the 16 logical qubits, instead of the 8+8 configuration, another choice is to use two nodes but with different capacities: we have one node hosting 4 logical qubits (node $A$), and another hosting 12 logical qubits (node $B$) (using 3 blocks of 4 logical qubits) - call this the 4+3x4 configuration. Now suppose we use one block of $[[36,4,6]]$ on $A$, and $B$ hosting 12 logical qubits, say using three $[[36,4,6]]$ blocks. 
Then, to perform an operation similar to the inter-node transversal CNOT, we could perform an inter-node transversal CNOT between the block on $A$ and an ancilla $[[36,4,6]]$ block on $B$, (followed by another similar inter-node transversal CNOT for uncomputation) (of course, one could argue that if we use ancilla qubits, then we don't need two nodes if we can use the ancilla qubits as computation qubits, but more generally, there could be other qubits in the first node not involved with the qubits in the second node). With the ancilla initialised to $\ket{\overline{0}}_L$, this would then effectively ``copy'' the logical qubits information across to node $B$, and then all related transversal operations can be carried out locally on $B$. 
An alternative to using $[[36,4,6]]BB$ blocks is to use $[[54,4,8]]BB$ blocks raising the number of required ebits but resulting in a larger distance.
Figure~\ref{fig:cf-trans-cnot-diffcap}  shows that the  transversal CNOT using $[[90,8,10]]$ (one block per node, i.e. 8+8 configuration, 8 logical qubits per node), despite using more ebits, still gives a lower LER compared to the transversal CNOT using $[[36,4,6]]$ (between the block in $A$ and the ancilla block in $B$) and the transversal CNOT using $[[54,4,8]]$; hence, the 8+8 configuration seems  better in terms of LER than the    4+12 configuration, in either 4 logical qubit encoding, though the $[[54,4,8]]$ comes closer (assuming local operations are similar in LER for both).  

The graph also shows the LER for an inter-node transversal CNOT involving two nodes, each node with a $[[144,12,12]]$ block (i.e., a 12+12 configuration). It  is noted that the number of physical ebits is 144, but yielding distance 12. So, comparing a 12+12 configuration  and a 8+16 configuration (i.e., a $[[90,8,10]]$ block on one node and two $[[90,8,10]]$ blocks on another node) (where a transversal CNOT will be between the $[[90,8,10]]$ block on one node and a  $[[90,8,10]]$ ancilla block on the other, and then, the physical CNOTs can be done locally), the 12+12 configuration can have a similar or even slightly lower LER even with an additional 144-90=54 ebits.  For the 8+16 configuration, however, one could use $[[120,8,12]]BB$ code blocks~\cite{rmy6-9n89}, with 24 fewer physical ebits than 144, yet achieving the same distance, which  could achieve a lower LER for a low enough PER (which is future work).  

{\bf Transversal GCZ.}
It is also interesting to note that with some BB codes, such as $[[120,8,12]]$, as constructed  in~\cite{2025arXiv251005211L}, which are self-dual,\footnote{There is the  the three-monomial  construction for $[[120,8,12]]$ in~\cite{rmy6-9n89}, but also the self-dual four-monomial construction for $[[120,8,12]]$ in~\cite{2025arXiv251005211L}.} there is a transversal $H$ operation, denoted by $\overline{H} \equiv H^{\otimes 120}$, acting as ${\overline{X}_L}_{2j-1} \leftrightarrow {\overline{Z}_L}_{2j}$ and ${\overline{Z}_L}_{2j-1}  \leftrightarrow {\overline{X}_L}_{2j}$, for $j \in \{1,2,3,4\}$, with respect to the logical $X$ and $Z$ operators, and so, we have $\overline{H} \not=\overline{H}_L^{\otimes 8}$, but have a multi-qubit transformation:
$\overline{H}= P_{\pi}  \overline{H}_L^{\otimes 8} $,
with $\pi = (1~2)(3~4)(5~6) (7~8)$, where the permutation operation $P_{\pi}$ can be a relabelling of the logical qubits. Now, for blocks $C$ and $T$, we have:
{\small
\begin{align*}
& (\overline{I} \otimes \overline{H}) {\overline{CNOT}}_{C,T} (\overline{I} \otimes \overline{H}) \\
& = (\overline{I}_L^{\otimes 8} \otimes P_{\pi} \overline{H}_L^{\otimes 8}) \bigotimes_{l=1}^{8} {\overline{CNOT}_L}[C^{(l)} \rightarrow T^{(l)}] (\overline{I}_L^{\otimes 8} \otimes P_{\pi} \overline{H}_L^{\otimes 8}) \\
& = P_{\pi}(T)~ \bigotimes_{l=1}^{8} {\overline{CZ}_L}[C^{(l)} \rightarrow T^{(l)}] ~P_{\pi}(T) 
\end{align*}
}where $P_{\pi}(T)$ applies the permutation on logical qubits in block $T$.  Also, we have:
\begin{align*}
& (\overline{I} \otimes \overline{H}) {\overline{CNOT}}_{C,T} (\overline{I} \otimes \overline{H}) \\
& = (I^{\otimes 120} \otimes H^{\otimes 120}) \bigotimes_{i=1}^{120} CNOT[{C_{i} \rightarrow T_{i}}] (I^{\otimes 120} \otimes H^{\otimes 120}) \\
& = \bigotimes_{i=1}^{120} CZ[{C_{i} \rightarrow T_{i}}] 
\end{align*}
Hence, we have 
$$
\bigotimes_{i=1}^{120} CZ[{C_{i} \rightarrow T_{i}}] = P_{\pi}(T)~ \bigotimes_{l=1}^{8} {\overline{CZ}_L}[C^{(l)} \rightarrow T^{(l)}] ~P_{\pi}(T)
$$
We then derive:
{\footnotesize
\begin{align*}
& \overline{GCZ}_{\{A,B,C,D\}} = \bigotimes_{i=1}^{120}GCZ_{\{A_i,B_i,C_i,D_i\}} \\
&  = \bigotimes_{i=1}^{120} \big(
CZ[A_{i} \rightarrow B_{i}]~ CZ[A_{i} \rightarrow C_{i}] ~ CZ[A_{i} \rightarrow D_{i}]  \\
& \quad \quad \quad \quad  CZ[B_{i} \rightarrow C_{i}]~ CZ[B_{i} \rightarrow D_{i}] ~CZ[C_{i} \rightarrow D_{i}] \big) \\
&  = \bigotimes_{i=1}^{120} CZ[A_{i} \rightarrow B_{i}]~ \bigotimes_{i=1}^{120} CZ[A_{i} \rightarrow C_{i}] ~\bigotimes_{i=1}^{120}  CZ[A_{i} \rightarrow D_{i}]  \\
&  \quad  \quad \bigotimes_{i=1}^{120}  CZ[B_{i} \rightarrow C_{i}]~\bigotimes_{i=1}^{120} CZ[B_{i} \rightarrow D_{i}] ~\bigotimes_{i=1}^{120}CZ[C_{i} \rightarrow D_{i}]  \\
&  =  P'_{\pi}~ \bigotimes_{l=1}^{8} \big(
{\overline{CZ}_L}[A^{(l)} \rightarrow B^{(l)}]~ {\overline{CZ}_L}[A^{(l)} \rightarrow C^{(l)}] ~ {\overline{CZ}_L}[A^{(l)} \rightarrow D^{(l)}]  \\
& \quad \quad \quad   {\overline{CZ}_L}[B^{(l)} \rightarrow C^{(l)}]~ {\overline{CZ}_L}[B^{(l)} \rightarrow D^{(l)}] ~{\overline{CZ}_L}[C^{(l)} \rightarrow D^{(l)}] \big)  ~P'_{\pi} \\
& = P'_{\pi}~  \bigotimes_{l=1}^{8}{\overline{GCZ}_L}_{\{A^{(l)},B^{(l)},C^{(l)},D^{(l)}\}} ~P'_{\pi}
\end{align*}
}where $P'_{\pi}=P_{\pi}(B)P_{\pi}(C)P_{\pi}(D)$. 
But note that this means we can compute the 8 logical $\overline{GCZ}_L$s concurrently via the transversal $\overline{GCZ}$, by adding a permutation of    the logical qubits on $A$:
\begin{align*}
&P_{\pi}(A) \bigotimes_{i=1}^{120}GCZ_{\{A_i,B_i,C_i,D_i\}}   P_{\pi}(A)  \\
&= P_{\pi}(A) P'_{\pi}~  \bigotimes_{l=1}^{8}{\overline{GCZ}_L}_{\{A^{(l)},B^{(l)},C^{(l)},D^{(l)}\}} ~P'_{\pi}  P_{\pi}(A)  \\
& \equiv  \bigotimes_{l=1}^{8}{\overline{GCZ}_L}_{\{A^{(l)},B^{(l)},C^{(l)},D^{(l)}\}}
\end{align*}
since $P_{\pi}(A)P'_{\pi} = P_{\pi}(A) P_{\pi}(B)P_{\pi}(C)P_{\pi}(D)$.

Note that this is not a single ${\overline{GCZ}_L}$ operation as we analyzed earlier. The transversal physical GCZ induces eight concurrent logical GCZ operations, up to the logical permutation $P_{\pi}$ associated with the transversal $H$ of the [[120,8,12]] encoding. Hence,  a choice of BB encoding with transversal $H$ can enable such computations, if needed.


\section{Discussion} 
\label{sec:dis}
We make the following observations:
\begin{itemize}
    \item If we want to reduce LER for a given computation, one possibility is to reduce the number of ebits, or reduce the amount of inter-node information transfer, and hence, a skewed configuration, such as 4+12, should work better than say 8+8.  But it seems that, in the cases studied here, the reduction in LER associated with the larger code distance can outweigh the additional errors associated with the larger number of ebits: we saw that 8+8 can have a lower LER than 4+12, and so, one can hypothesize that a 12+12 can have a lower LER than 8+16 (at least for the inter-node transversal CNOT gadget using ancilla blocks above). Hence,
 from the above, reducing ebits between nodes is not necessarily a dominant factor in deciding the choice of encoding and nodes to use - a larger physical qubit encoding block could yield operations/circuits with lower LERs via higher distance even if requiring a larger number of inter-node ebits.
 \item By a similar argument as above, for the same $k$, we could use a higher $n$ even at the cost of more physical ebits required - e.g., use $[[54,4,8]]$ instead of $[[36,4,6]]$ and use $[[120,8,12]]$ instead of $[[90,8,10]]$. 
 \item  There is a question of whether to use one large block  or multiple smaller blocks, for all the logical qubits on a node - large blocks can yield more efficient encodings (efficiency in the sense of higher $kd^2/n$) and higher distances yielding even lower LERs for transversal CNOTs, even if a typical transversal CNOT might require many more ebits (and ancilla qubits) if done in ``one step'' - as we have seen, the 8+8 configuration has lower LER than the 2x4+2x4 configuration even with $p_{ebit}=50p$.
\item With larger blocks, 
 operations are coarser (in the sense of involving many qubits in ``one step'' of an operation, e.g., a transversal CNOT with $n$ qubits performs $k$ logical CNOTs at the same time, and larger blocks with larger $k$ yields larger number of such concurrent operations) - such concurrency can be useful in some distributed computations requiring it (e.g., for the global gates as shown here). But sparsely connected circuits where only certain qubits of one node interacts with certain other qubits on another node might benefit from smaller code blocks (while trading-off distance), e.g., since some operations might  involve only selected blocks from each node.
 \item In the configurations examined here, using larger blocks and fewer nodes can reduce the number of inter-node operations and, for the noise models studied, can result in lower LERs. Whether this remains true for other circuits requires further study.
\end{itemize}
However, the above are only based on a limited study. One could compare LERs of much larger circuits with different node configurations and different encoding blocks. Also, we used  BB codes only here with transversal operations, and only a limited set - e.g., others from~\cite{2025arXiv251005211L,rmy6-9n89} can be explored, and other codes such as surface codes. We
 have not explored the use of conversions, e.g., from a $[[144,12,12]]$ code block to three $[[54,4,8]]$ blocks and conversely, whether this would be useful for distributed computations - we have assumed uniform encodings for all blocks involved in inter-node transversal CNOTs.

\section{Conclusion}
\label{sec:conc}
We have provided a study of encoding and node choices for inter-node transversal CNOTs, and while our examples are mainly indicative, we conclude the following:
\begin{itemize}
\item Rather than choosing the encoding and node configuration based on ebit consumption alone, or $n$, $d$, or $k$ alone, one can consider efficiency or their combined  effect on the LER of the actual distributed circuit.   
\item Given a distributed quantum circuit and a set of heterogeneous QPUs, encoding and node selection can be formulated as a constrained optimization problem.
\begin{equation*}
\boxed{
\begin{aligned}
\text{Select }(P,E)
&=
\underset{(P,E)\in\mathcal{C}_{\rm feasible}}{\arg\min}
\operatorname{LER}_{\rm circuit}(P,E),
\end{aligned}
}
\end{equation*}
where $P$ is the partition/allocation of logical qubits across nodes, and $E$ is the choice of QEC encodings and block decomposition for those nodes.
The appropriate encoding and node configuration is therefore circuit- and architecture-dependent.
 
\end{itemize}

\section*{Acknowledgements}
The author acknowledges the use of ChatGPT-5.6 Luna to assist in 
(i) generating (by generalizing from smaller hand-constructed examples followed by manual fixes) Figures~\ref{fig:gcz120}, \ref{fig:non-local-part-gcz-2nodes} and \ref{fig:three-cnot-blocks},  and (ii) generating the Python code for plotting the graphs in Figures~\ref{fig:cf-trans-cnot} and \ref{fig:cf-trans-cnot-diffcap}. The author verified all algorithmic logic and executed the code, and takes full responsibility for the  results.
\bibliographystyle{plain}
\bibliography{refs}

\section*{Appendix~A}
\label{sec:appA}
\begin{figure*}[t]
\centering
  \includegraphics[width=1.0\textwidth, height=9cm]{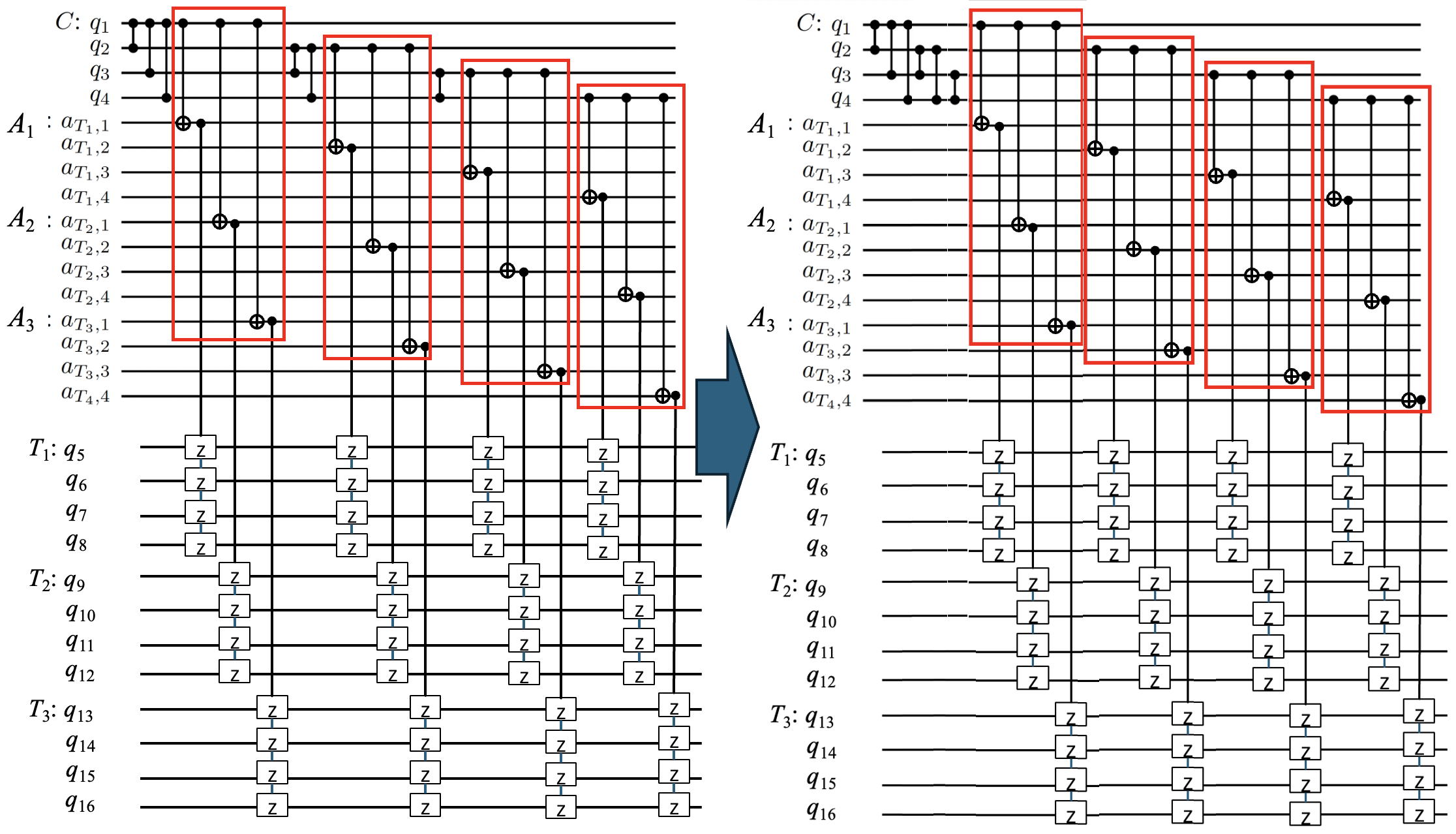} 
    \includegraphics[width=1.0\textwidth, height=9cm]{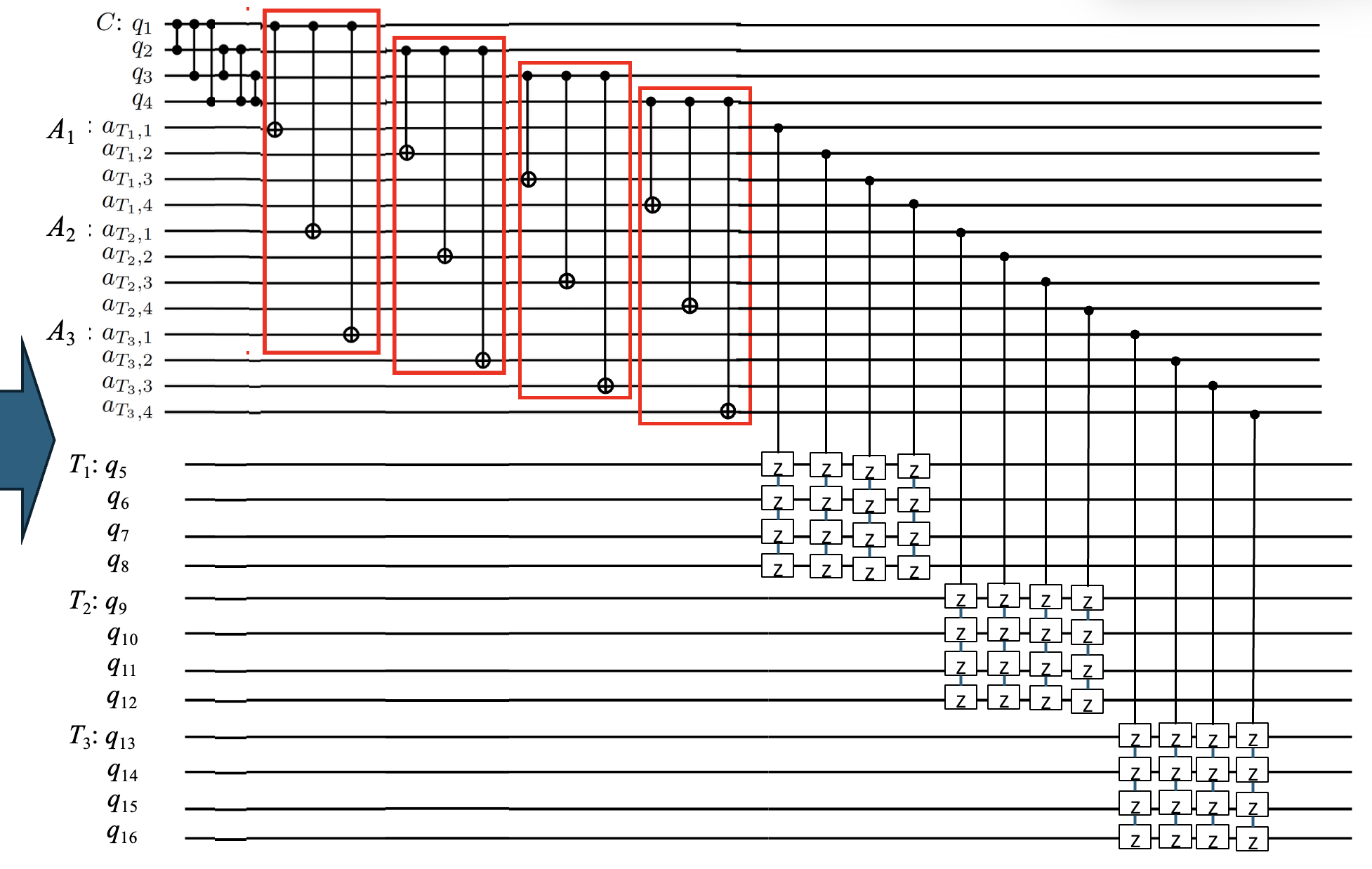} 
  \caption{Adapted  from~\cite{transversalfanout}, partial distributed implementation of the 54 $\overline{CZ}_L$s from  Figure~\ref{fig:gcz54} using ancilla qubits, showing (i) reordering with the local $\overline{CZ}_L$s ``batching'' and preceding  the rest of the inter-node operations; (ii) the  inter-node operations (marked in the four  red boxes) can be reordered and implemented via transversal CNOTs shown in Figure~\ref{fig:three-cnot-blocks}. With the ancilla initialised to $\ket{\overline{0}}_L$, we effectively ``copy'' the logical qubits information from one node across to another node, and then all related $\overline{CZ}_L$ operations can be carried out locally. Uncomputation of the ancilla blocks is  needed via another sequence of transversal CNOTs and not shown here.}
  \label{fig:dgcz54}
\end{figure*}

The operations marked by red boxes in Figure~\ref{fig:dgcz54} can be rearranged into transversal CNOTs in Figure~\ref{fig:three-cnot-blocks}.

\end{document}